\documentstyle[12pt]{article}
\begin{document}
\begin{titlepage}
\begin{flushright}
TPI-MINN 98/27-T \\
UMN-TH-1731/98   \\
hep-th/9812157\\
\end{flushright}

\vspace{0.8cm}

\begin{center}
\Large{\bf{A Quasi-Exactly Solvable $N$-Body Problem \\
with the $sl(N+1)$ Algebraic
Structure}} ~\\
\end{center}
\vspace{5mm}
\begin{center}
{\bf Xinrui Hou  and  M. Shifman} \\
\vspace{5mm}    
Theoretical Physics Institute \\
University of Minnesota \\
Minneapolis, MN 55455
\end{center}
\vspace{4cm}

\begin{abstract}
Starting from a one-particle quasi-exactly solvable system, 
which is characterized by an intrinsic $sl(2)$ algebraic structure and   
the energy-reflection symmetry, we  construct a daughter $N$-body  
Hamiltonian presenting a  deformation of the Calogero model. The 
features of  this Hamiltonian are (i) it reduces to a quadratic 
combination of the  generators of $sl(N+1)$; (ii) the interaction potential 
contains  two-body terms and interaction with the force center at the 
origin; (iii)  for quantized values of a certain cohomology 
parameter $n$ it is quasi-exactly  solvable, the multiplicity of states 
in the algebraic sector is $(N+n)!/(N!n!)$; (iv) the energy-reflection 
symmetry of the parent system is preserved.
\end{abstract}
\end{titlepage}

\section{Motivation and results}  
\label{sec1}

A hidden Lie-algebraic structure of certain Hamiltonians allows one 
to  generate quasi-exactly solvable (QES) spectral 
problems~\cite{a1}--\cite{a4} with quantized cohomology parameters, 
determining a part  of the spectrum which is tractable algebraically. 
Several  $N$-body  problems of this type were suggested 
recently~\cite{a6}--\cite{a5}. The most instructive for our purposes 
is the model constructed in Ref.~\cite{a5}.  It presents an $sl(2)$-based 
deformation of the Calogero model~\cite{a7}, which was, in turn, shown 
to be  Lie-algebraic in the fundamental work~\cite{a8}. Here we report a 
quasi-exactly solvable  $N$-body problem obtained as an $sl(N+1)$ 
deformation of the  Calogero model.

\vspace{0.1cm}

The Hamiltonian of the model has the form
\begin{eqnarray}
{\cal H}&\!\!\!=\!\!\!&-\sum_{i=1}^{N}\frac{\partial^{2}}{\partial
x_{i}^{2}} +\sum_{i=1}^{N}\left\{-[4n+4\alpha(N-1)+4\beta+3]x_{i}^{2} 
+x_{i}^{6}+\frac{g}
 {4x_{i}^{2}}\right\}    \nonumber   \\ [0.3cm]
 & &+ g\sum_{\stackrel{i,j=1}{i\neq j}}^{N}\left\{\frac{1}{(x_{i}-
x_{j})^{2}} + \frac{1}
 {(x_{i}+x_{j})^{2}}\right\}  \,,
\label{b1}
\end{eqnarray}
where $g$ is a positive constant, the constants $\alpha$ and $\beta$ 
are  defined as
\begin{equation}
\alpha=\frac{\sqrt{1+4g}+1}{2}\,,\qquad \beta=\frac{\sqrt{1+g}+1}{4} 
\,,   
\label{b2}
\end{equation}
and $n$ is the quantized cohomology parameter taking integer 
values, $n=0,1,2,...$.
The number $\nu$ of states in the algebraic sector is related to $n$ 
as follows:
\begin{equation}
\nu=\frac{(N+n)!}{N!n!} \,.
\label{b3} 
\end{equation}
The first line in Eq.~(\ref{b1}) presents the kinetic term plus a
self-interaction while the second line contains a two-body coupling plus
``mirror-reflected" terms. In Sec. \ref{sec2} this Hamiltonian will  be 
shown  to be reducible to  a quadratic combination of the  generators of 
$sl(N+1)$. In Sec.  \ref{sec3}
we consider a specific example, while Sec.~\ref{sec4} is devoted to 
generalizations. Note that, omitting the $x^2$ and $x^6$ terms in the
first  line,  one arrives at a particular case of the so-called $BC_n$ model.
The latter was shown to be exactly-solvable, its Lie-algebraic 
representation is based on $sl(N+1)$~\cite{a14}. Thus, the quasi-exactly 
solvable system (\ref{b1}) we present, can be also viewed as a 
deformation of the  $BC_n$ model. In fact,  key technical elements
of the proof  are essentialy the same as  those worked out  in 
Refs.~\cite{a5,a8,a14}.

\vspace{0.1cm}

As well-known (see~\cite{a9}),  the original Calogero model can be 
obtained from the  one-dimentional harmonic oscillator,
\begin{equation}
H_{0}=-\frac{1}{2}\frac{d^{2}}{dx^{2}}+\frac{1}{2}x^{2} \,.
 \label{b4}
\end{equation}
To this end  one observes that the eigenfunctions of the spectral 
problem~(\ref{b4}) have the form
\begin{equation}
\psi_{n}(x)=P_{n}(x)\, e^{-x^2/2}   \,,  
\label{b5}
\end{equation} 
where $P_{n}(x)$ is a polynomial of degree $n$. It can be always 
parametrized  by the position of its zeros,
\begin{equation}
P_{n}(x)\propto (x-x_{1})\cdots (x-x_{n}) \,.
 \label{b6}
\end{equation}
A generic time-dependent wave function
\begin{equation}
\Psi(t,x)=\sum_{n}C_{n}\psi_{n}(x)\, e^{-iE_{n}t}  
 \label{b7}
\end{equation}
can be written as 
\begin{equation}
\Psi(t,x) \propto (x-x_{1}(t))\cdots (x-x_{n}(t))\, e^{-x^2/2}\,,
\label{b8}
\end{equation}
where $x_{i}(t),(i=1,\ldots, n)$ are time-dependent functions.

\vspace{0.1cm}

One can check~\cite{a9} that the equations of motion for $x_{i}(t)$ 
are those following from the Lagrangian
\begin{equation}
{\cal L}_{\rm daughter}=\frac{1}{2}\sum_{i=1}^{N}\dot{x}_{i}^{2}-
\frac{1}{2}\left\{ \sum_{i=1}^{N}
x_{i}^{2}+g\sum_{\stackrel{i,j=1}{i\neq j}}^{N}\frac{1}{(x_{i}-
x_{j})^{2}}\right\} 
 \label{b9}
\end{equation}
at $g=1$. This is a particular case of the original Calogero 
system~\cite{a7}. Since the energy eigenvalues of the harmonic 
oscillator are commensurate, it is quite obvious that the 
system (\ref{b9}) has infinitely many periodic (complex)
classical trajectories, which is indicative of the exact solvability. 
Indeed, as well-known, the quantum Calogero model is exactly 
solvable~\cite{a7,a8}.

\vspace{0.1cm}

Recently it was suggested~\cite{a10} to apply a  similar strategy to 
generate a deformed Calogero model  from the 
one-dimensional  polynomial QES system with the sextic interaction.
Starting from a sextic QES potential of a general form, a deformed 
classical Calogero Lagrangian was obtained, with the quadratic, 
quartic  and sextic self-interaction terms, see Eq. (26) in~\cite{a10}. A 
somewhat simplified form of this Lagrangian which will be useful for 
our purposes (see below) is
\begin{equation}
{\cal L}_{\rm daughter}=\frac{1}{2}\sum_{i=1}^{2N}\dot{x}_{i}^{2}-
\frac{1}{2}
\sum_{i=1}^{2N}\left\{-2Ax_{i}^{2}+x_{i}^{6}\right\}    
-\frac{1}{2}\sum_{\stackrel{i,j=1}{j\neq i}}^{2N}\frac{1}
 {(x_{i}-x_{j})^{2}} \,,
\label{b10}
\end{equation}
where $A$ is a numerical constant.

\vspace{0.1cm}

We use, as a starting point, a special case of the polynomial QES 
system possessing the energy-reflection (ER) 
symmetry~\cite{a11,a12}:
\begin{equation}
H=\frac{1}{2}\left[-\frac{d^{2}}{dx^{2}}+x^{6}-(8j+3)x^{2}\right]  \,.
\label{b11}
\end{equation}
If the parameter $j$ is chosen to be 
\begin{displaymath}
j=\frac{N}{2}\,,     
\end{displaymath}
the wave functions belonging to the algebraic sector have the form
(see~\cite{a11}) 
\begin{equation}
\psi_{n}=P_{2N}(x)\, e^{-x^{4}/4}  
  \label{b12}
\end{equation} 
where 
\begin{equation}
P_{2N}(x)\propto (x-x_{1})\cdots(x-x_{N})(x-x_{N+1})\cdots(x-
x_{2N}) \,,
   \label{b13}
\end{equation}
with the additional condition
\begin{equation}
x_{N+i}=-x_{i},\qquad i=1,\ldots, N  \, .
     \label{b14}
\end{equation}
Since $N=2j$ and $j$ is half-integer, the parameter $N$ is integer,
of course.  We stress that the polynomial in Eq. (\ref{b13}) is of 
degree $2N$ rather than $N$. 

\vspace{0.1cm}

By considering linear combinations of the type~(\ref{b7}) we convert 
the parameters $x_{i}$ into time-dependent functions. The classical 
daughter Lagrangian,  from which the classical equations of motion for 
$x_{i}(t)$ ensue, is that of Eq.~(\ref{b10}).

\vspace{0.1cm}

Why we require the parent system to have the ER symmetry? In the 
general case neither of the trajectories $x_{i}(t)$ obtained 
in~\cite{a10} is  periodic. This is due to the fact that the energy 
eigenvalues are not commensurate.  If, however, the parent system is ER 
symmetric, as in Eq.~(\ref{b11}),  then one can consider special 
combinations (\ref{b7}) of the type
\begin{displaymath}
\Psi(t,x)=A\, \psi_{E}(x)\, e^{-iEt}+B\, \psi_{-E}(x)\, e^{iEt}   
\end{displaymath}
where $\{\psi_{E},\psi_{-E}\}$ is any of the ER pairs of the 
system~(\ref{b11}); there are $[(N+1)/2]$ such pairs (here the square 
brackets denote the integer part).

\vspace{0.1cm}

It is obvious that this choice gives rise to a periodic (complex) 
trajectory for $x_{i}(t)$. The number of such trajectories in the daughter 
system  following from (\ref{b7}) is $[(N+1)/2]$. (Compare this with the 
harmonic  oscillator and the daughter Calogero model where the 
number of periodic  trajectories is infinite.) This gives us a hint that a 
daughter system obtained from~(\ref{b11}) may be quasi-exactly 
solvable.

\vspace{0.1cm}
 
To build the quasi-exactly solvable system based on $sl(N+1)$  we 
impose the constraints~(\ref{b14}). More exactly, we allow for 
a general set of the coupling constants preserving the structure 
of~(\ref{b10}).  The coefficients in front of $\partial^{2}/\partial 
x_{i}^{2}$ and $x_{i}^{6}$ can be always made equal to unity by an 
overall  rescaling of variables and the Hamiltonian. The coefficient in 
front of the two-body coupling term is set to $g$ where $g$ is an 
{\em arbitrary positive} constant (the  positivity is needed for the 
stability of the problem). The coefficients in front of
$x_{i}^{2}$ and $x_{i}^{-2}$ are then unambiguously fixed by the 
requirement of the quasi-exact solvability. In this way we arrive at 
the  Hamiltonian~(\ref{b1}). It is worth  emphasizing that the constraint
(\ref{b14}) is absolutely crucial, the number of degrees of freedom in 
the Hamiltonian~(\ref{b1}) is twice smaller than the number of 
zeroes in Eq.~(\ref{b13}), and, hence, twice smaller than the number of 
degrees of freedom in (\ref{b10}).

\section{Proof of algebraization}
\label{sec2}

To uncover the hidden algebraic structure of the Hamiltonian 
(\ref{b1}) we, first, perform the change of variables,
\begin{equation}
x_{i}\rightarrow \xi_{i}\equiv x_{i}^{2} \,.
    \label{b15}     
\end{equation} 
In this way we arrive at 
\begin{eqnarray}
{\cal H}_{\xi}&=&-
\sum_{i=1}^{N}\left(4\xi_{i}\frac{\partial^{2}}{\partial\xi_{i}^{2}}+2
\frac{\partial}
{\partial\xi_{i}}\right)   \nonumber  \\[0.2cm]
 & &+\sum_{i=1}^{N}\left\{-[4n+4\alpha (N-
1)+4\beta+3]\xi_{i}+\xi_{i}^{3}+\frac{g}
{4\xi_{i}} \right\}   \nonumber    \\[0.2cm]
 & &+2g\sum_{\stackrel{i,j=1}{i\neq j}}^{N}
\frac{\xi_{i}+\xi_{j}}{(\xi_{i}-\xi_{j})^{2}}\,.     
      \label{b16} 
\end{eqnarray}
The next step of the program (for a review see~\cite{a11}) is a 
gauge transformation (sometimes referred to as similarity 
transformation)
\begin{equation}
{\cal H}_{\xi}\rightarrow {\cal H}_{G}=e^{a}{\cal H}_{\xi}e^{-a} \,,  
\label{b17}
\end{equation}
where the phase $a$ appropriate to the Hamiltonian (\ref{b16}) is
\begin{equation}
a=\ln\Psi_{0},\qquad 
\Psi_{0}=\prod_{\stackrel{i,j=1}{i<j}}^{N}(\xi_{i}-
\xi_{j})^{-\alpha}\,
\prod_{i=1}^{N}
\xi_{i}^{-
\beta}\,\exp\left\{\frac{1}{4}\sum_{i=1}^{N}\xi_{i}^{2}\right\} \,.  
\label{b18}
\end{equation}
On physical grounds it is clear that the constants $\alpha$ and 
$\beta$ must  be positive. In fact, as will be seen shortly, these 
constants are the  positive roots of the equations
\begin{equation}
\alpha^{2}-\alpha=g \, ,\qquad 
4\beta^{2}-2\beta=\frac{1}{4}g \,  ,  
\label{b19}
\end{equation}
see Eq.~(\ref{b2}). The parametrization of the phase $a$ above 
generalizes that exploited in the earlier  proofs of the
algebraization  of the Calogero  and $BC_n$ models. The
transformation~(\ref{b17}) is  non-unitary. The spectra of ${\cal
H}_{\xi}$ and ${\cal H}_{G}$ are  the same, however, while the
eigenfunctions  are related as follows:
\begin{displaymath}
\psi_{\xi}=\psi_{G}e^{-a}\, . 
\end{displaymath}
Under the gauge transformation~(\ref{b17})
\begin{equation}
\frac{\partial}{\partial \xi_{i}}\rightarrow
e^{a}\frac{\partial}{\partial \xi_{i}}e^{-a}=\frac{\partial}{\partial
\xi_{i}}+\sum_{\stackrel{j=1}{ j\neq i}}^{N}\frac{\alpha}{\xi_{i}-
\xi_{j}}
+ \frac{\beta}{\xi_{i}} -\frac{1}{2}\xi_{i} \, .
  \label{b20}
\end{equation}
The corresponding transformation of $\partial^{2}/\partial\xi_{i}^{2}$ 
follows from~(\ref{b20}) with the use of the identity
\begin{equation}
\sum_{\stackrel{i,j,k=1}{i\neq j\neq k}}^{N}\frac{\xi_{i}}
{(\xi_{i}-\xi_{j})(\xi_{i}-\xi_{k})}=0 \,.
\label{b21}
\end{equation}
It is not difficult to get that
\begin{eqnarray}
{\cal H}_{G}
&=&-
\sum_{i=1}^{N}\left\{4\xi_{i}\frac{\partial^{2}}{\partial\xi_{i}^{2}}
+(8\beta+2)\frac{\partial}{\partial\xi_{i}}-4\xi_{i}^{2}\frac{\partial}
{\partial\xi_{i}}+4n\xi_{i}\right\}   \nonumber   \\[0.2cm]
 & &-4\alpha \sum_{\stackrel{i,j=1}{i\neq j}}^{N}\frac{1}{\xi_{i}-
\xi_{j}}\left(\xi_{i}
\frac{\partial}{\partial\xi_{i}}-
\xi_{j}\frac{\partial}{\partial\xi_{j}}\right) \,. 
   \label{b22} 
\end{eqnarray}
The terms proportional to $\xi_{i}^{3},\,\,\, \xi_{i}^{-1}$ and 
$(\xi_{i}-\xi_{j})^{-2}$ in Eq.~(\ref{b16}) cancel after the gauge 
transformation. The cancellation is ensured by the 
condition~(\ref{b19}).

\vspace{0.1cm}

Now we are just one step short of revealing  the $sl(N+1)$ algebraic 
structure of the Hamiltonian~(\ref{b1}). Following~\cite{a8,a14}, we 
pass from  the original set of variables  $\{\xi_{i},\ldots,\xi_{N}\}$ to a
new set  $\{\tau_{1},\ldots,\tau_{N}\}$ where $\tau_{i}'s$ are the 
elementary symmetric polynomials,
\begin{equation}
\tau_{1}=\sum_{i=1}^{N}\xi_{i} ,\qquad
\tau_{2}=\sum_{\stackrel{i,j=1}{i<j}}^{N}\xi_{i}\xi_{j},\quad  
\ldots,\quad     
\tau_{N}=\xi_{1}\xi_{2}\ldots\xi_{N} \,.
 \label{b23}                     
\end{equation} 
It is convenient to define, additionally, $\tau_{0}=1$ and 
$\tau_{j}=0$  if $j<0$ or $j>N$. A collection of formulae helpful in 
rewriting ${\cal H}_{G}$ in terms of $\tau_{i}$'s is given in Appendix. 
Applying these formulae it is not difficult to obtain
\begin{eqnarray}
{\cal H}_{G}&\!\!\!=\!\!\!&-
4\sum_{k=1}^{N}\sum_{l=1}^{N}\left[\sum_{j=1}^{k}(l-
k+2j-1)
\tau_{k-j}\tau_{l+j-
1}\right]\frac{\partial^{2}}{\partial\tau_{l}\partial\tau_{k}} 
\nonumber     \\[0.2cm]
 & &-4\alpha\sum_
{k=1}^{N}\left(N-k+1\right)\left(N-
k\right)\tau_{k-1}\frac{\partial}{\partial
\tau_{k}} 
-(8\beta+2)\sum_{k=1}^{N}\left(N-
k+1\right)\tau_{k-1}\frac{\partial}
{\partial\tau_{k}} \nonumber   \\[0.2cm]
 & &+4\sum_{k=1}^{N}\left[\tau_{1}\tau_{k}-
(k+1)\tau_{k+1}\right]\frac{\partial}{\partial\tau_{k}}  
-4n\tau_{1} \,. 
  \label{b24}    
\end{eqnarray}

\vspace{0.2cm}

What remains to be done is to rewrite this expression as a quadratic
combination of the generators of the  $sl(N+1)$ algebra. The 
differential realizations of the generators of $sl(N)$ are
thoroughly studied in the mathematical literature. We will not dwell 
on details referring the reader to the textbooks \cite{tb}. The explicit 
parametrization 
appropriate for our purposes can be  found e.g. in~\cite{a8,a14} (a 
simple heuristic derivation is presented in~\cite{a13}).  It is 
built on $N$ variables and refers  to a degenerate representation -- the
symmetrized product of $n$  fundamental representations. The 
$N^{2}+2N$ generators of $sl(N+1)$ in this  realization  take the 
form~\footnote{The number of the generators listed in Eq.~(\ref{b25}) is 
$N^2+2N+1$. However, one linear combination, ${\cal N}_0 +\sum_i{\cal 
N}_i$,  reduces to a constant and is, therefore, redundant. It is easy to 
see that
$N$ independent generators from the set $\{{\cal N}_0\, , \,\,\,{\cal
N}_i\}$ form the Cartan subalgebra of the  $sl(N+1)$ algebra.}
\begin{eqnarray}
{\cal D}_{i}&=&\frac{\partial}{\partial\tau_{i}}\,,\qquad  
i=1,2,\ldots,N\, ,   
\nonumber   \\[0.2cm]
{\cal U}_{i}&=&\tau_{i}{\cal N}_{0}\,,\qquad  i=1,2,\ldots,N\,, 
\nonumber \\ [0.2cm]
{\cal N}_{0}&=&n-\sum_{i=1}^{N}\tau_{i}\frac{\partial}
{\partial\tau_{i}}\,,  \nonumber    \\[0.2cm]
{\cal N}_{i,j}&=&\tau_{i}\frac{\partial}{\partial\tau_{j}}\,,\qquad 
i\neq 
j,\,\,\, i,j=
1,2,\ldots,N \,,  \nonumber    \\[0.2cm]
{\cal N}_{i}&=&\tau_{i}\frac{\partial}{\partial\tau_{i}}\,, \qquad  
i=1,2,\ldots,N\,,    
 \label{b25}
\end{eqnarray}
where the parameter $n$ is an integer, the number of the 
fundamental  representations in the symmetrized product. One can 
explicitly verify  that the algebra of the operators in~(\ref{b25}) is that 
of  $sl(N+1)$.  Moreover, for integer values of
$n$ the algebra of the generators (\ref{b25}) has a finite-dimensional
representation of dimension $\nu=(N+n)!/(N!n!)$ 
composed of the homogeneous products of $\tau_{i}$ of degree
$1,2,\ldots,n$.

\vspace{0.1cm}

After some simple algebraic transformations  the 
Hamiltonian~(\ref{b24}) reduces to a quadratic combination of the 
generators, namely
\begin{eqnarray}
{\cal H}_{G}&\!\!\!=\!\!\!&-
4\sum_{k=1}^{N}\sum_{l=1}^{N}\sum_{j=1}^{k}
(l-k+2j-1)({\cal N}_{k-j,l}{\cal N}_{l+j-1,k}-\delta_{j1}{\cal N}_{k-
1,k})  \nonumber  \\[0.2cm]
 &
&-4\alpha\sum_{k=1}^{N}\left(N-k+1\right)
\left(N-k\right){\cal N}_{k-1,k}-
(8\beta+2)\sum_{k=1}^{N}\left(N-k+1
\right){\cal N}_{k-1,k}  
\nonumber  \\[0.2cm]
 & &-4\sum_{k=1}^{N}(k+1){\cal N}_{k+1,k} -4{\cal U}_{1}\,.
\label{b26} 
\end{eqnarray} 
Note that here we define
\begin{displaymath}
{\cal N}_{0,k}={\cal D}_{k}\,,   \qquad    k=1,2,\ldots,N,
\end{displaymath}
for convenience.  We will stick to  this definition in 
what follows.

\vspace{0.1cm}

The representation (\ref{b26}) proves the quasi-exact solvability of the 
$N$-body  problem (\ref{b1}). The number of levels in the algebraic 
sector is  given by Eq. (\ref{b3}).

\vspace{0.1cm}

Following the arguments of Ref.~\cite{a12}, it is easy to verify that all
levels in the algebraic sector are split in pairs under the
energy-reflection symmetry. The negative-energy state  goes into 
the positive-energy state and {\it vice versa} under the transformation
$x_k\to i x_k$, much in the same way as in the parent system 
(\ref{b11}).
  
\section{Example}
\label{sec3}

To get a more transparent idea of the system we built, it is 
instructive  to consider a specific example. Let us consider the two-body 
problem,
\begin{equation}
N=2,\;   \;   n=1 \,.
\label{b27}
\end{equation}   
The underlying algebra is $sl(3)$;  at $n=1$, the number of levels 
in the algebraic sector is $\nu=3$. For simplicity, we put $g=1$. 
Then
\begin{displaymath}
\alpha=\frac{\sqrt{5}+1}{2},\;   \;    \beta=\frac{\sqrt{2}+1}{4}\,.
\end{displaymath}
The dynamical variables are $x_{1}$ and $x_{2}$ (or $\xi_{1}$ and
$\xi_{2}$.) One introduces $\tau_{1}$ and $\tau_{2}$ according to
Eq.(\ref{b23}). The algebraic representation of the (gauge 
transformed) Hamiltonian is 
\begin{eqnarray}
{\cal H}_{G}&\!\!\!=\!\!\!&-
4\tau_{1}\frac{\partial^{2}}{\partial\tau_{1}^{2}}-
16\tau_{2}
\frac{\partial^{2}}{\partial\tau_{1}\partial\tau_{2}}-
4\tau_{2}\tau_{1}\frac
{\partial^{2}}{\partial\tau_{2}^{2}}-
8\alpha\frac{\partial}{\partial\tau_{1}}
-    2(8\beta+2)\frac{\partial}{\partial\tau_{1}}\nonumber    
\\[0.2cm]
 & &-(8\beta+2)\tau_{1}
\frac{\partial}{\partial\tau_{2}}+4(\tau_{1}^{2}-2\tau_{2})
\frac{\partial}{\partial\tau_{1}}   
+4\tau_{2}\tau_{1}\frac{\partial}{\partial\tau_{2}}-4\tau_{1}  
 \nonumber    \\[0.3cm]
&\!\!\!=\!\!\!&-4{\cal N}_{1}}{{\cal D}_{1}-16{\cal N}_{2,1}{\cal 
D}_{2}
-4{\cal N}_{2}{\cal N}_{1,2}-8\alpha {\cal D}_{1}  \nonumber    
\\[0.3cm]
 & &-2(8\beta+2){\cal D}_{1}-(8\beta+2){\cal N}_{1,2}-8{\cal N}_{2,1}
-4{\cal U}_{1}    \,.
\label{b28}    
\end{eqnarray}
The algebraic sector consists of three levels. The corresponding
eigenvalues and eigenfunctions are
\begin{eqnarray}
E_{\pm}&\!\!\!=\!\!\!&\pm\sqrt{32(4\beta+\alpha+1)} \,,   
\nonumber   \\[0.2cm]
\psi_{\pm}&\!\!\!=\!\!\!&\left( 1 \mp 
\frac{\sqrt{2(\alpha+1+4\beta)}}{4\beta+2\alpha+1}
(x_{1}^{2}+x_{2}^{2})+\frac{2}{4\beta+2\alpha+1}\, x_{1}^{2}
x_{2}^{2}\right)     \nonumber \\[0.2cm]
 & &\times \left( x_{1}^{2}-x_{2}^{2}\right)^{\alpha} \, 
x_{1}^{2\beta}\, x_{2}^{2\beta}\, 
\exp\left\{-\frac{ x_{1}^{4}+x_{2}^{4}}{4}\right\}  \,,
\label{b29}  
\end{eqnarray}
and
\begin{eqnarray}
E&\!\!\!=\!\!\!&0\,,   \nonumber   \\[0.2cm]
\psi&\!\!\!=\!\!\!&\left(1-\frac{2}{4\beta+1}\, 
x_{1}^{2}x_{2}^{2}\right)
\left( x_{1}^{2}-x_{2}^{2}\right)^{\alpha} \, x_{1}^{2\beta}\, 
x_{2}^{2\beta}\, \exp\left\{-\frac{ x_{1}^{4}+x_{2}^{4}}{4}\right\} 
 \,.
 \label{b30}   
\end{eqnarray}
The normalization constants in the wave functions above are omitted. 
We have worked out a few more examples with $n>1$. The 
corresponding formulae, although perfectly explicit and closed, are 
quite cumbersome, and we  do not reproduce them here.

\section{Generalizations}
\label{sec4}

The Hamiltonian (\ref{b1}) admits various generalizations. The most 
obvious extension is the inclusion of the quartic terms in the 
potential
\begin{eqnarray}
{\cal H}&\!\!\!=\!\!\!&-\sum_{i=1}^{N}
\frac{\partial^{2}}{\partial x_{i}^{2}}
  \nonumber \\[0.2cm] 
 & &+\sum_{i=1}^{N}\left\{\left( b^{2}-a[4n+4\alpha(N-1) +
4\beta+3]\right) x_{i}^{2} +
 a^{2}x_{i}^{6}+2abx_{i}^{4}+\frac{g}
 {4x_{i}^{2}}\right\}    \nonumber \\[0.2cm]  
 & &+ g\sum_{\stackrel{i,j=1}{i\neq
j}}^{N}\left\{\frac{1}{(x_{i}-x_{j})^{2}} + 
\frac{1}{(x_{i}+x_{j})^{2}}\right\}\,,
 \label{b31}
\end{eqnarray}
where $a,b$ are constants,  with $a>0$ and $\alpha,\beta$ are the 
same as in Eq.~(\ref{b19}). Then, after the change of variables 
(\ref{b15}),
\begin{eqnarray}
{\cal H}_{\xi}&\!\!\!=\!\!\!&
-\sum_{i=1}^{N}\left( 4\xi_{i}\frac{\partial^{2}}{\partial\xi_{i}^{2}}+
2\frac{\partial}
{\partial\xi_{i}}\right)   \nonumber  \\[0.2cm] 
 & &+\sum_{i=1}^{N}\left\{\left( b^{2}-a[4n+4\alpha (N-
1)+4\beta+3]\right)
\xi_{i}+a^{2}\xi_{i}^{3}+
 2ab\xi_{i}^{2}+\frac{g}{4\xi_{i}} \right\}  \nonumber     \\[0.2cm] 
 & &+2g\sum_{\stackrel{i,j=1}{i\neq j}}^{N}
\frac{\xi_{i}+\xi_{j}}{(\xi_{i}-\xi_{j})^{2}} \,.
\label{b32}     
\end{eqnarray}
Next,  we make the gauge transformation~ (\ref{b17}) with the phase 
similar to~(\ref{b18}), namely
\begin{displaymath}
\Psi_{0}=\prod_{\stackrel{i,j=1}{i<j}}^{N}(\xi_{i}-\xi{_j})^{-\alpha}\,
\prod_{i=1}^{N}
\xi_{i}^{-\beta}\,
\exp\left\{\frac{a}{4}\sum_{i=1}^{N}\xi_{i}^{2}+\frac{b}{2}
\sum_{i=1}^{N}\xi_{i}\right\}\, .
\end{displaymath}
As a result, the Hamiltonian takes the form of a quadratic 
combination of the  generators of $sl(N+1)$, 
\begin{eqnarray}
{\cal H}_{G}&\!\!\!=\!\!\!&-4\sum_{k=1}^{N}\sum_{l=1}^{N}
\sum_{j=1}^{k}(l-k+2j-1)({\cal N}_{k-j,l}{\cal N}_{l+j-1,k}-
\delta_{j1}{\cal N}_{k-1,k})  \nonumber   \\[0.2cm] 
 & &-4\alpha\sum_{k=1}^{N}(N-k+1)(N-k){\cal N}
_{k-1,k}    
 -(8\beta+2)\sum_{k=1}^{N}(N-k+1){\cal N}_{k-1,k}
-4a{\cal U}_{1}   \nonumber     \\[0.2cm] 
 & &-4a\sum_{k=1}^{N}(k+1){\cal N}_{k+1,k}
+4b\sum_{k=1}^{N}k{\cal N}_{k}   +2b\alpha N(N-1)+(4b\beta+b)N \,.  
\label{b33}
\end{eqnarray}
This representation makes  the quasi-exact solvability of (\ref{b31}) 
obvious.  Needless to say that with the quartic terms added the 
energy-reflection symmetry is  lost, generally speaking. 

\vspace{0.1cm}

Another generalization of the model~(\ref{b1}) is 
supersymmetrization along the lines of~\cite{a14}.
The sufficient condition for a straightforward supersymmetrization is
the following representation of the potential:
\begin{equation}
V(x_i) = \sum_{i=1}^N \left(\frac{\partial W}{\partial x_i}\right)^2\,.
\label{b34}
\end{equation}
One can check that the potential in Eq.~(\ref{b1}) does have the 
required form provided $n=0$. In this case the superpotential $W$ is
\begin{equation}
W(x_i) =-\frac{1}{4}\sum_{i=1}^{N}x_{i}^{4}+\alpha\sum_{\stackrel{i,j=1}
{i<j}}^{N}\ln (x_{i}^{2}-x_{j}^{2})+\beta \sum_{i=1}^{N}\ln x_{i}^{2}\, .
\label{b35}
\end{equation}
Adding the fermion terms  following the standard prescription, we 
get a  Hamiltonian with the minimal supersymmetry. It describes a 
system of $N$ particles with spin, with a certain two-body 
interaction (including the spin interaction). Since we were forced to 
put $n=0$, only the ground state wave function is known 
algebraically. Whether one can built algebraically the wave functions 
of the excited states remains to be seen. Another open question is 
whether one can built non-minimal supersymmetric models at 
$n\neq 0$. 

\section*{Acknowledgments}

Useful discussions and correspondence with A. Turbiner and A. 
Ushveridze are gratefully acknowledged. 
This work was supported in part by DOE under the grant number 
DE-FG02-94ER40823.

\newpage

\section*{Appendix : Transition $\xi_{i}\rightarrow \tau_{j}$} 
\renewcommand{\theequation}{A.\arabic{equation}}
\setcounter{equation}{0}

The variable $\tau_{i}$ (the elementary symmetrical polynomials) 
are defined in Eq. (\ref{b23}). The transition $\xi_{i}\rightarrow 
\tau_{j}$ can be  carried out by virtue of the following basic 
identities:
\begin{eqnarray}
& & \tau_{k-1}-\xi_{i}\frac{\partial\tau_{k-
1}}{\partial\xi_{i}}=\frac{\partial
\tau_{k}}{\partial\xi_{i}}  \,, \\[0.3cm] 
& & \sum_{i=1}^{N}\xi_{i}\frac{\partial\tau_{k}}{\partial\xi_{i}}=k
\tau_{k},\qquad k=1,2,\ldots, N\,,  \\[0.3cm] 
& & \sum_{\stackrel{i,j=1}{i\neq j}}^{N}\frac{\partial^{2}\tau_{k}}
{\partial\xi_{i}\partial\xi_{j}}\, \xi_{i}\xi_{j}=k(k-1)\tau_{k}  \,.
\end{eqnarray}
Exploiting these identities, after some straightforward but tedious 
algebraic manipulations, one gets
\begin{eqnarray}
\sum_{i=1}^{N}\frac{\partial}{\partial\xi_{i}}&\!\!\! =\!\!\! &
\sum_{k=1}^{N}[N\tau_{k-1}-(k-1)\tau_{k-1}]\frac{\partial}
 {\partial\tau_{k}}  \,,  
\\[0.3cm] 
\sum_{i=1}^{N}\xi_{i}^{2}\frac{\partial}{\partial\xi_{i}}
&\!\!\! =\!\!\! &
\sum_{k=1}^{N}
[\tau_{1}\tau_{k}-(k+1)\tau_{k+1}]\frac{\partial}
{\partial\tau_{k}}     
 \,,  
\\[0.3cm] 
\sum_{i=1}^{N}\xi_{i}\frac{\partial^{2}}{\partial\xi_{i}^{2}}
&\!\!\! =\!\!\! &
\sum_{j=1}^{k}(l-k+2j-1)\tau_{k-j}\tau_{l+j-1}    
 \,,  
\end{eqnarray}
\begin{equation}
\sum_{\stackrel{i,j=1}{i\neq j}}^{N}
\frac{1}{\xi_{i}-\xi_{j}}\, \left(\xi_{i}\frac{\partial}{\partial\xi_{i}}-
\xi_{j}\frac{\partial}{\partial\xi_{j}}\right)\,
 =
\sum_{k=1}^{N}(N-k+1)(N-k)\tau_{k-1}\frac{\partial}
 {\partial\tau_{k}}\,.
\end{equation}

\end{document}